\documentclass[%
 reprint,
 amsmath,amssymb,
 aps,
 pra,
]{revtex4-2}

\usepackage{flushend}
\usepackage{verbatim}
\usepackage{verbatim,upgreek}
\usepackage{chngcntr}
\usepackage{graphicx}
\usepackage{dcolumn}
\usepackage{bm}
\usepackage{comment}
\usepackage{amsmath}
\usepackage{physics}
\usepackage{hyperref}
\hypersetup{
    colorlinks=true,
    citecolor=blue,
    }

\usepackage{color}

\begin{document}

\preprint{APS/123-QED}

\title{Cavity-enhanced solid-state nuclear spin gyroscope}

\author{Hanfeng Wang$^{1}$}
\email{hanfengw@mit.edu}
\author{Shuang Wu$^{1,2}$}

\author{Kurt Jacobs$^{3,4}$}

\author{Yuqin Duan$^{1}$}

\author{Dirk R. Englund$^{1}$}
 
\author{Matthew E. Trusheim$^{1,3}$}
 \email{matthew.e.trusheim.civ@army.mil}

\affiliation{
$^{1}$ Massachusetts Institute of Technology, 50 Vassar Street, Cambridge, MA 02139, USA\\
$^{2}$ Honda Research Institute USA, Inc., San Jose, CA 95134, USA\\
$^{3}$ U.S. Army DEVCOM Army Research Laboratory, Adelphi, MD 20783, USA\\
$^{4}$ Department of Physics, University of Massachusetts Boston, Boston, MA 02125, USA}

\begin{abstract}

Solid-state quantum sensors based on ensembles of nitrogen-vacancy (NV) centers in diamond have emerged as powerful tools for precise sensing applications. Nuclear spin sensors are particularly well-suited for applications requiring long coherence times, such as inertial sensing, but remain underexplored due to control complexity and limited optical readout efficiency. In this work, we propose cooperative cavity quantum electrodynamic (cQED) coupling to achieve efficient nuclear spin readout. Unlike previous cQED methods used to enhance electron spin readout, here we employ two-field interference in the NV hyperfine subspace to directly probe the nuclear spin transitions. We model the nuclear spin NV-cQED system (nNV-cQED) and observe several distinct regimes, including electromagnetically induced transparency, masing without inversion, and oscillatory behavior.
We then evaluate the nNV-cQED system as an inertial sensor, indicating a rotation sensitivity improved by three orders of magnitude compared to previous solid-state spin demonstrations. Furthermore, we show that the NV electron spin can be simultaneously used as a comagnetometer, and the four crystallographic axes of NVs can be employed for vector resolution in a single nNV-cQED system. These results showcase the applications of two-field interference using the nNV-cQED platform, providing critical insights into the manipulation and control of quantum states in hybrid NV systems and unlocking new possibilities for high-performance quantum sensing.

\end{abstract}

\maketitle

\textit{Introduction - }Nitrogen-vacancy (NV) centers in diamond have emerged as a promising platform for a variety of sensing modalities \cite{barry2020sensitivity,degen2017quantum,doherty2013nitrogen,kim2023nanophotonic,wang2023field} due to favorable attributes including room-temperature spin polarization and readout, atomic-scale size~\cite{du2017control,pelliccione2016scanned,hu2024developing}, and long coherence times~\cite{bauch2018ultralong, stanwix2010coherence,wang2021observation}. NV sensors generally use magnetic resonance spectroscopy of the electron ground-state spin triplet transition frequencies. The nuclear spin degree of freedom is less explored as a sensor, since it is not directly accessible for optical polarization and readout, and has potentially disadvantageous properties such as reduced resonant frequency and gyromagnetic ratio. However, reduced sensitivity to noisy magnetic fields results in a long coherence time \cite{kalb2018dephasing}, which is advantageous for quantum memory and inertial sensing applications \cite{jarmola2021demonstration,wang2024hyperfine,wang2024emulated}.

\begin{figure}
\includegraphics[width = 0.48\textwidth]{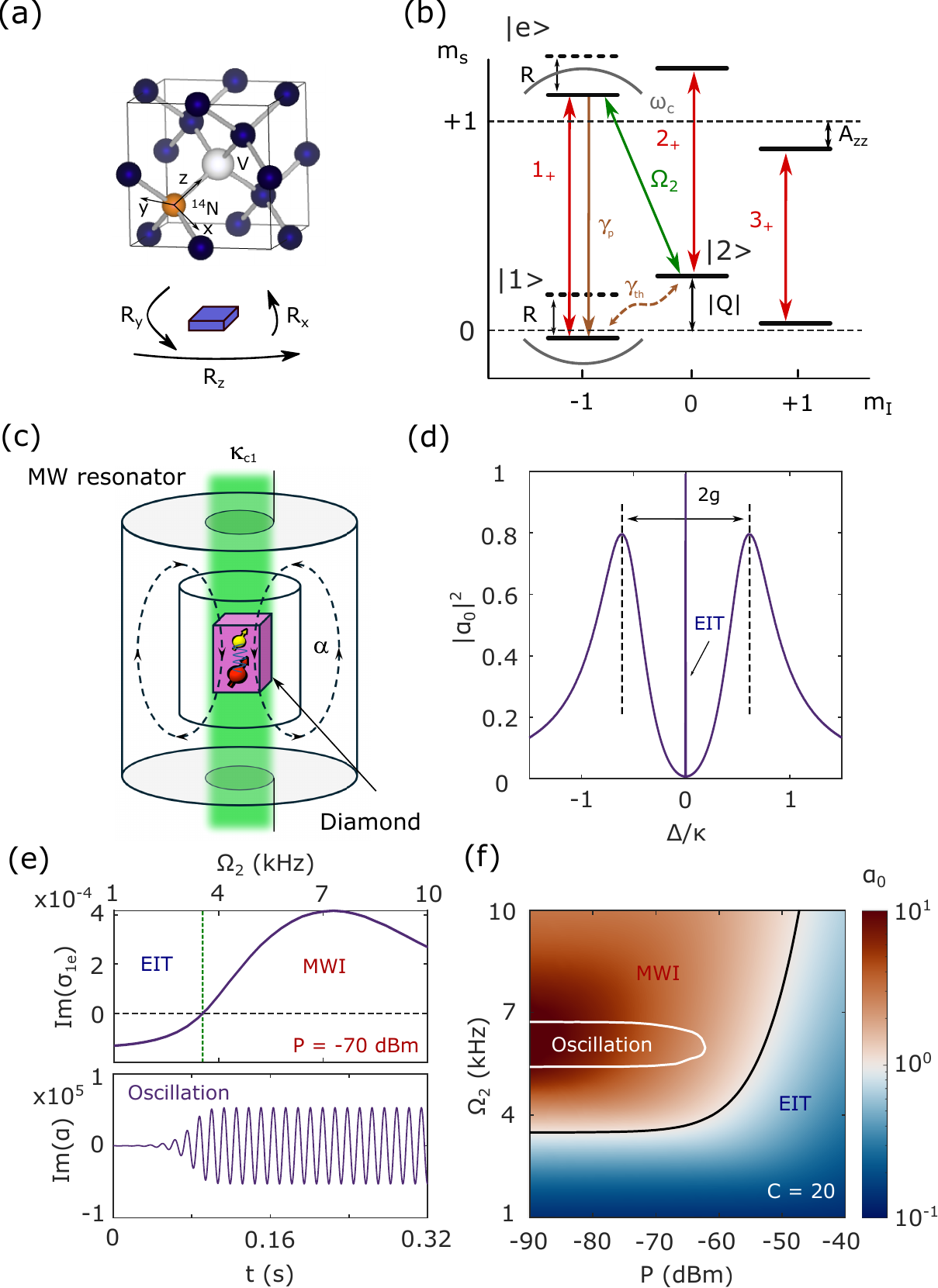}
\caption{\label{fig:1} \textbf{Two-field interference in nNV-cQED.} (a) Top: NV crystal structure. Bottom: diamond with NVs rotates with rate $\mathbf{R} = \{R_x,~R_y,~R_z\}$. (b) NV energy level structure. The transition $|1\rangle \leftrightarrow |e\rangle$ is coupled to the cavity mode for the cavity-enhanced readout. A driving field $\Omega_2$ is applied between the spin-exchanging transition $|2\rangle \leftrightarrow |e\rangle$. (c) Hybrid system with a microwave resonator and an NV spin ensemble. A green laser is applied to continuously polarize the NV spin to the $|m_s=0\rangle$, and a detection loop is incorporated to measure the reflection signal from the resonator. (d) $|\alpha_0|^2$ as a function of detuning $\Delta/\kappa$ within strong coupling regime. An EIT feature appears around the resonant frequency. (e) Top: Im($\sigma_{1e}$) as a function of $\Omega_2$. The EIT (MWI) regime features a negative (positive) Im($\sigma_{1e}$). Bottom: Time dynamics of Im($\alpha$). (f) $\alpha_0$ as a function of $P$ and $\Omega_2$. The solid line indicates the perfect EIT condition. The boundary of the oscillation regime is marked as a white line. }
\end{figure}

Nuclear spin gyroscopes promise stable long-term performance with competitive short-term sensitivity. Devices based on atomic vapors have achieved inertial angle random walk (ARW) comparable with the optical sensors at the mdeg/$\sqrt{\mathrm{h}}$ level, but corresponding performance has not been achieved using nuclei in the solid state \cite{stockton2011absolute}. The concept of using an NV ensemble as a solid-state nuclear spin gyroscope \cite{ledbetter2012gyroscopes,maclaurin2012measurable,ajoy2012stable} describes a standard quantum limit (SQL) as $\eta =  1/\sqrt{NT_2^*}$, where $T_2^*$ is the nuclear spin dephasing time, and $N$ is the number of spins. This leads to a quantum-limited ARW of 0.1 mdeg/s/$\sqrt{\mathrm{Hz}}$ for a spin ensemble of $N=10^{14}$ and $T_2^*=2$ ms, which outperforms similarly-sized (mm-scale) solid-state systems such as microelectromechanical systems and fiber optic gyroscopes \cite{lai2020earth,hokmabadi2019non}. However, current state-of-the-art NV-based nuclear spin gyroscopes achieve an ARW of $\eta_{\mathrm{exp}}$ = 4.7 deg/s/$\mathrm{\sqrt{Hz}}$ \cite{jarmola2021demonstration,soshenko2021nuclear}. This ARW remains significantly higher than the SQL predicted in the theoretical outline \cite{ajoy2012stable}, and can be quantified as an inverse readout fidelity $\sigma_n=\eta_{\mathrm{exp}}/\eta \sim 10^4$ that captures all measurement imperfections. This leaves room for dramatic improvement towards the SQL.

NV-cavity quantum electrodynamic (cQED) coupling is a promising avenue that has been applied to enhance readout fidelity for electron spin ensembles \cite{wang2024spin,eisenach2021cavity}. However, the manipulation of nuclear spin has not been demonstrated with a cQED scheme, as nuclei are not directly spin-polarizable using optics and have low gyromagnetic ratio and resonance frequency that substantially reduces achievable spin-cavity coupling. In this work, we extend our electron spin-based cQED scheme to realize a nuclear spin readout (nNV-cQED) using two-field interference. Like electromagnetically induced transparency (EIT) gyroscopes in atomic vapor systems, this mechanism exhibits a linewidth limited by the nuclear spin coherence, providing an exceptional sensitivity to rotation. 
We then show that the electron spin coherence of the cavity-coupled transition becomes positive, leading to a cavity gain and enhanced sensitivity in the masing without inversion (MWI) regime \cite{zhu1992lasing}. Optimizing over performance and additional spin refrigeration effects \cite{fahey2023steady}, we indicate an ARW of 1.5 mdeg/s/$\sqrt{\mathrm{Hz}}$ with an inverse readout fidelity of $\sigma_n\sim 37$, surpassing the sensitivity of state-of-the-art NV gyroscopes by three orders of magnitude.

To further enhance functionality, we introduce a ``4-EIT" scheme enabling vector inertial sensing and electron spin comagnetometry \cite{soshenko2021nuclear}. The vector resolution is realized via four NV axes, which differs from atomic gyroscopes where external magnetic bias primarily determines nuclei orientation. An electron spin comagnetometer is implemented by measuring common mode and difference-frequency shifts of driving fields, improving bias stability by compensating for noise due to electron spin and temperature. Unlike previous cavity readout approaches based on collective polariton measurements \cite{wang2024spin}, each transition (rotation and magnetic field) in the 4-EIT scheme can be measured independently with minimal crosstalk or performance degradation.
The nNV-cQED system enhances inertial sensitivity while providing a pathway for high-precision, room-temperature atomic clocks. Its versatile design opens the door to broader applications in quantum information processing and exploration of novel quantum phenomena in hybrid solid-state systems.

\textit{Two-field interference} - The nuclear spin gyroscope relies on two-field interference in a hybrid nNV-cQED system, where a probe microwave field's state is controlled by a driving field via coherent NV spin interaction. Two-field interference phenomena, such as EIT, have been demonstrated for NVs and atomic systems \cite{huillery2021coherent,jamonneau2016coherent,vanier2005atomic,gray1978coherent}. However, EIT in highly cooperative cavity-coupled ensembles is rarely reported, and NV ensembles are further complicated by optical polarization and inhomogeneous broadening.

We consider an NV ensemble in a bulk diamond, as illustrated in Fig. 1(a). The NV ground state comprises an $S = 1$ electronic triplet state coupled via hyperfine interaction to the $I = 1~^{14}N$ nuclear states. Taking the \{111\} as the $z$-axis, the NV Hamiltonian reads \cite{huillery2021coherent}:
\begin{equation}    
\mathcal{H}_{\mathrm{NV}} / h=D \hat{S}_z^2+\gamma_e \hat{\mathbf{S}} \cdot \mathbf{B}+Q \hat{I}_z^2-\gamma_n \hat{\mathbf{I}} \cdot \mathbf{B}+\hat{\mathbf{S}} \cdot \underline{\underline{\mathcal{A}}} \cdot \hat{\mathbf{I}}
\end{equation}
where $D=2.87~\mathrm{GHz}$ and $Q=-4.945~\mathrm{MHz}$ are the electron and nuclear spin zero-field splittings, $\gamma_e=2\pi\times 2.8~\mathrm{MHz} / \mathrm{G}$ and $\gamma_n=2\pi\times0.308~ \mathrm{kHz} / \mathrm{G}$ are the electron and nuclear spin gyromagnetic ratios, and $\mathcal{A}$ is the diagonal hyperfine interaction tensor. The NV ground-state level structure based on Eq. (1) is shown in Fig. 1(b). An off-axis magnetic field $B$ mixes the electron and nuclear spin states, enabling forbidden spin-exchange transitions. We focus on a $\Lambda$-system formed by electron-nuclear spin states $\{|0,-1\rangle,~|0,0\rangle,~|1,-1\rangle\}$, and label these as $\{|1\rangle, ~|2\rangle,~|e\rangle\}$ respectively.

We consider the interaction of the NV system with a quantized cavity mode field $a$ coupled to the transition $|1\rangle\leftrightarrow|e\rangle$, driven by an external probe field $J$, as shown in Fig. 1(c). The Hamiltonian of the coupled NV-cavity system reads:
\begin{equation}
\begin{aligned}
\mathcal{H}&=\Delta a^{\dagger} a +\Delta_{s} \sigma_{ee}+\Delta_{2}\sigma_{22}+\sum g_s(a \sigma_{e 1}+a^{\dagger} \sigma_{1 e})\\
&+\Omega_2\left(\sigma_{2 e}+\sigma_{e 2}\right)+J(a^{\dagger}+a)
\end{aligned}
\end{equation}
Here $\Delta=\omega_c-\omega_d$ is the cavity detuning between detection frequency $\omega_d$ and cavity frequency $\omega_c$, $\Delta_{s}=\omega_{s}-\omega_d$ is the detuning between spin transition $|1\rangle \leftrightarrow |e\rangle$ frequency $\omega_{s}$ and detection frequency, and $\Delta_2 = \omega_2+|Q|-\omega_d$ is the detuning between the driving field and the probe field. The rate $g_s$ ($g$) corresponds to single (total) coupling strength, $J$ to the probe field strength, and $\Omega_2$ to the Rabi frequency of the spin-flip driving field.

We use the quantum master equation to solve the dynamics for nuclear spin manipulation based on the Hamiltonian in Eq. (2), together with loss processes using a Lindbladian open system formalism \cite{wang2024spin}. The cavity mode has a relaxation rate $\kappa = \kappa_{\mathrm{c}} + \kappa_{\mathrm{c}1}$, where $\kappa_{\mathrm{c}}$ is intrinsic and $\kappa_{\mathrm{c}1}$ is coupling to a microwave probe line. The NV ensemble includes several polarization and decoherence processes: optical polarization driving spins to ground states at rate $\gamma_\mathrm{p}$; electron spin decoherence at rate $\Gamma$; thermal depolarization at rate $\gamma_{\mathrm{th}}$; and nuclear spin linewidth $\Gamma_n$. All processes are assumed homogeneous in the following discussions.

\begin{figure}
\includegraphics[width = 0.49\textwidth]{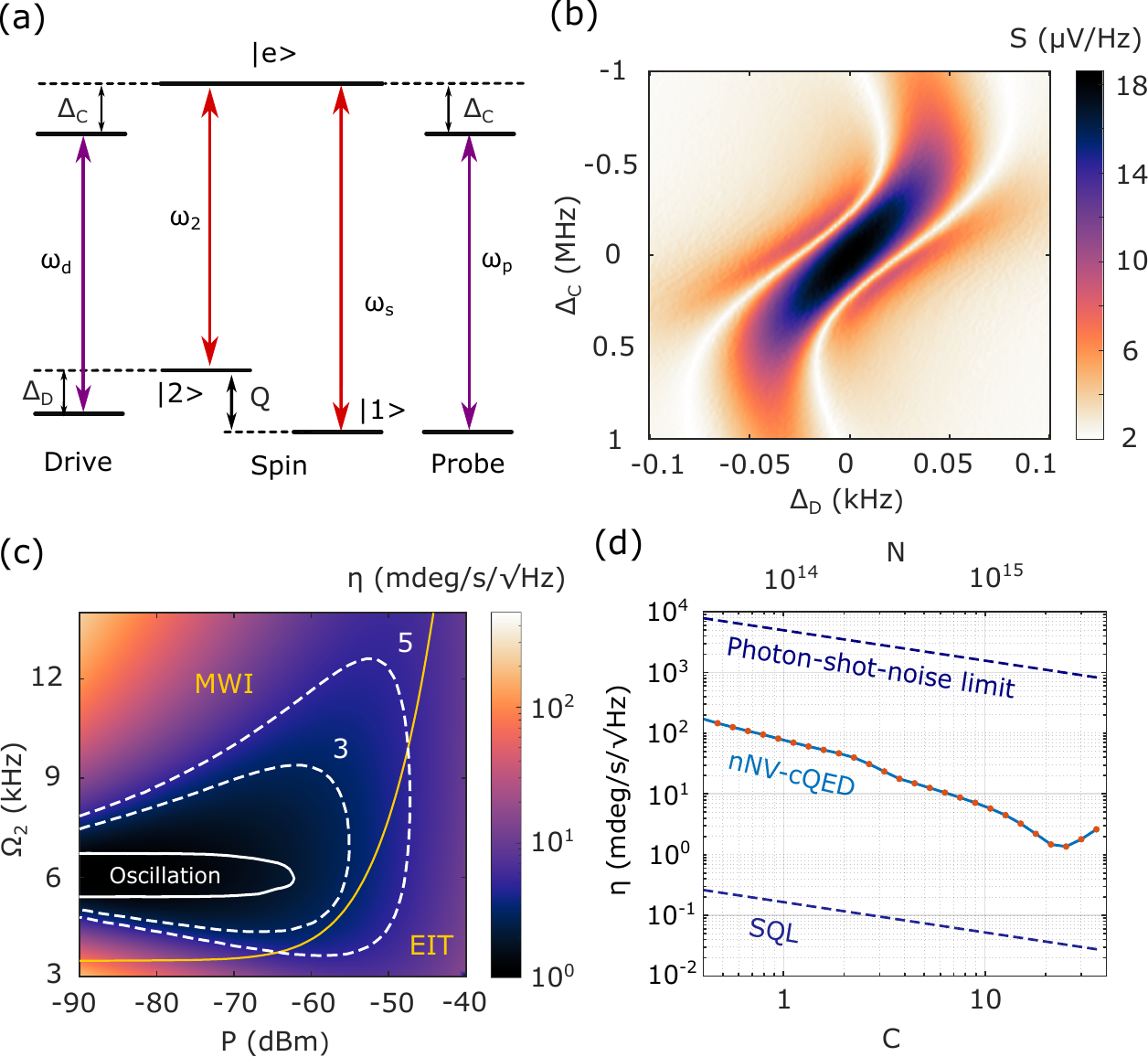}
\caption{\label{fig:1} \textbf{nNV-cQED gyroscope.} (a) Level structure for rotation sensing. The signal can be divided into common-mode signal $\Delta_\mathrm{C}$ and difference-mode signal $\Delta_\mathrm{D}$. (b) $S$ as a function of $\Delta_\mathrm{C}$ and $\Delta_\mathrm{D}$. (c) ARW as a function of probe power $P$ and driving field strength $\Omega_2$. The boundary of the oscillatory regime is marked as a solid white line. An optimal ARW of 1.5 mdeg/s/$\sqrt{\mathrm{Hz}}$ can be achieved. (d) Sensitivity as a function of cooperativity. We plot the photon-shot-noise limit and SQL as a comparison. The best inverse readout fidelity is around $\sigma_n\sim 37$ at $C\sim20$.}
\end{figure}

The primary parameter of interest for nNV-cQED operation is the cavity field, modulated by the spin ensemble and driving field. We first consider the case where the spin and cavity are on resonance with achievable system cooperativity $C\sim 20$ \cite{Supplementary}, and the system is driven with a rate $\Omega_2 = 6$ kHz on resonance with the spin-exchanging transition while probed with power $P=-55$ dBm. The solution for the cavity field as a function of input probe detuning is shown in Fig.\ 1(d), using an input-normalized cavity field $|\alpha_0|^2 \equiv |\kappa_{c1}\alpha/J|^2$. In this strong coupling regime, the hybridization of the cavity mode with the spin ensemble results in a Rabi splitting of the cavity field maxima, characteristic of the collectively-coupled system. A narrow EIT feature appears at the two-photon resonance condition where $\Delta = 0$ due to two-field interference. The EIT linewidth and amplitude are determined by the interplay of driving and probe fields with nuclear spin decoherence. The minimum linewidth is set by the nuclear-spin coherent dark state's lifetime, and the EIT contrast compared to the cavity-coupled spin system scales with system cooperativity.

We now consider the nNV-cQED device under different two-field driving conditions. In the steady-state, the spin ensemble depends on the cavity field via the relation $\alpha = (J-ig_sN\sigma_{1e})/(\kappa/2)$, where the spin coherence $\sigma_{1e}$ depends on the spin-exchange driving field, $\Omega_2$, as $\sigma_{1e} = (-ig_s\alpha - i\Omega\sigma_{12})/(\Gamma/2)$ (valid when most population is in $\ket{1}$). The relation between $\Omega_2$ and $\sigma_{1e}$ is shown in Fig. 1(e), top. Low driving fields result in a negative Im($\sigma_{1e}$) dominated by the cavity field $\alpha$, corresponding to imperfect EIT. As the drive $\Omega_2$ increases, ground state coherence $\sigma_{12}$ is built corresponding to coherent population trapping of a nuclear spin dark state, and the term $i\Omega\sigma_{12}$ compensates the resonant cavity probe field $ig_s\alpha$. This increases $\sigma_{1e}$ and reduces the phase shift of the cavity field due to spins, with $\sigma_{1e} = 0$ corresponding to complete cancellation (perfect EIT). Interestingly, as $\Omega_2$ continues to increase, $\sigma_{1e}$ becomes positive, resulting in gain for the cavity field (MWI). This regime produces the largest two-field interference visibilities and is, therefore, the most relevant for sensor operation. In the limit $\Omega_2\sigma_{12} \gg g_s\alpha$, the cavity field has $\alpha \propto \frac{g_sN\Omega_2}{\kappa\Gamma} = \frac{\Omega_2}{g_s}C$, where $C$ is the collective cooperativity. Increased cooperativity then results in increased MWI gain for fixed driving field.

These regimes are illustrated in Fig. 1(f), which shows the normalized steady-state cavity field $\alpha_0$ under varying two-field drive parameters. As probe power increases, the system demonstrates saturation effects. In the low-power regime, increased probe power does not destroy the perfect EIT condition as coherence $\sigma_{12}$ compensates increased $g_s\alpha$. As $P \rightarrow -55$ dBm and beyond, however, increased $\Omega_2$ is required to maintain perfect EIT due to $\sigma_{12}$ saturation. This combination induces Autler-Townes splitting in the dressed state picture, broadening the resonant peak and decreasing EIT contrast \cite{cohen1996autler}.

In this high-cooperativity device, an oscillatory regime exists where $ig_sN\sigma_{1e}$ exceeds the probe field $J$, providing sustaining gain that overcomes cavity losses. Compared with the EIT and MWI regimes, where output frequency locks with the probe field, frequency in the oscillating regime is determined by spin-cavity parameters and the driving field. In Fig. 1(e) bottom, we plot Im($\alpha$) as a function of time for detuning $\Delta_b = 2\pi\times 80$ Hz in the oscillating regime, showing a beat note with frequency $\Delta_b$ compared to the probe field. We note that sustained oscillation without inversion has rarely been reported in literature \cite{zibrov1995experimental,mompart2000lasing}.

\textit{Gyroscope performance - }Having shown the features of the nNV-cQED system under two-field drive, we now consider performance in an inertial sensing application. The nuclear spin behavior under a rotation $\mathbf{R}$, shown in Fig. 1(a,b), is described by the Hamiltonian $\mathcal{H}_R = \mathbf{R}\cdot \mathbf{I}$ \cite{soshenko2021nuclear}.
We divide the sensing signal into common mode ($\Delta_\mathrm{C}$) and difference mode ($\Delta_\mathrm{D}$), as shown in Fig. 2(a). The common mode describes the sensing signal shifting the state $|e\rangle$ while keeping nuclear spin energies constant. The difference mode describes the sensing signal shifting $|2\rangle$ while changing nuclear energy difference from $Q$ to $Q+\Delta_\mathrm{D}$. A rotation $R$ results in both common- and difference-mode shifts equal to the rotation rate, but the resonant linewidth for the difference mode is limited by nuclear spin linewidth ($\sim$ 80 Hz), three orders smaller than the electron spin linewidth-limited common signal ($\sim$ 0.33 MHz).

\begin{figure}
\includegraphics[width = 0.5\textwidth]{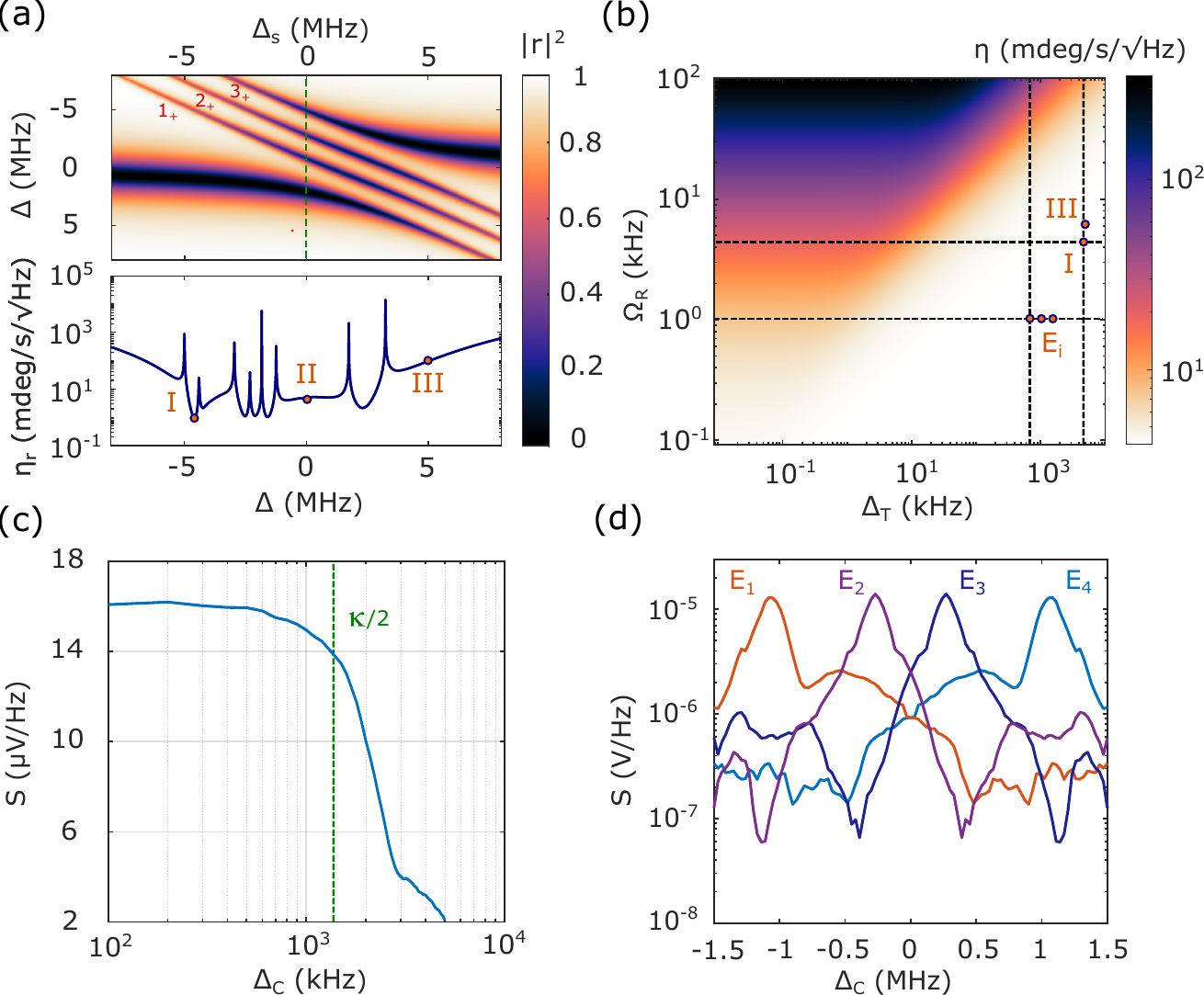}
\caption{\label{fig:1} \textbf{Comagnetometry and vector gyroscope.} (a) Electron spin-based comagnetometer. Top: The reflection $|r|^2$ as a function of $\Delta$ and $\Delta_s$. Bottom: The linecut for the sensitivity along $\Delta = 0$ for $P_2=-40$ dBm. The numerals indicate the frequencies of I: Comagnetometer probe field. II: EIT probe field. III: EIT driving field. (b) $\eta$ in the presence of an additional field with Rabi rate $\Omega_R$ and detuning $\Delta_T$. The red dots indicate field combinations for co-magnetometry and vector operation. (c) Signal $S$ as a function of $\Delta_\mathrm{C}$. $S$ drops significantly beyond the cavity linewidth. (d) The response $S$ for subensembles $E_i$ as a function of $\Delta_\mathrm{C}$.  }
\end{figure}

The device's sensitivity is determined by the signal and noise when probing the system: $\eta = \mathcal{L}/S$. The signal per unit frequency shift is defined as $S=\frac{V\partial\mathrm{Im(r)}}{\partial\Delta_\mathrm{D}}$, where $V$ is the probe field input voltage and $r=-1+\alpha_0$ is the probe reflection coefficient. We plot $S$ as a function of $\Delta_{\mathrm{C}}$ and $\Delta_{\mathrm{D}}$ in Fig. 2(b) with the same spin-cavity parameters as in the previous section. The optimal $S$ is 18 $\mu$V/Hz with a linear dynamical range of 20 Hz.

The noise performance is typically limited by Johnson-Nyquist noise at room temperature $\mathcal{L}_{\mathrm{JN}}$. However, as shown in previous works \cite{wang2024spin,fahey2023steady}, the NV ensemble can act as a spin refrigerator, performing below the Johnson-Nyquist limit even with active microwave probes. The noise performance can then be expressed as $\mathcal{L}=\xi\mathcal{L}_{\mathrm{JN}}$, with the refrigeration prefactor $\xi=0.45$ \cite{Supplementary}.

We plot the sensitivity $\eta$ as a function of $\Omega_2$ and $P$ in Fig. 2(c). An optimal sensitivity of 1.5 mdeg/s/$\sqrt{\mathrm{Hz}}$ is achievable in the MWI regime, while in the EIT regime the best sensitivity is 3.5 mdeg/s/$\sqrt{\mathrm{Hz}}$. In the oscillation regime, the field frequency is determined by spin-cavity parameters and the driving field. The sensitivity of such a frequency modulation system depends on oscillator phase noise \cite{barry2023ferrimagnetic}, which is outside the scope of our discussion. The rotation SQL of this system is 0.04 mdeg/s/$\sqrt{\mathrm{Hz}}$, and the optimum sensitivity corresponds to an inverse readout fidelity of $\sigma_n=37$, three orders better than photon-shot-noise-limited sensitivity \cite{jarmola2021demonstration}.

Finally, we analyze the ARW dependence on nNV-cQED system cooperativity. In Fig. 2(d), we plot the ARW dependence with $C$, given the same single-spin coupling strength and loaded quality factor. Here cooperativity is tuned by changing the ensemble size. In the low-$C$ regime ($C<10$), the ARW improves linearly with $C$. The cooperativity is then decreased super-linearly in the regime of $10<C<24$, attributed to MWI. After $C>24$, the maximum-sensitivity operating point enters the oscillating regime and degrades performance. The best readout fidelity for our system is $\sigma_n = 37$ with $C = 20$ and ensemble size $2.4\times 10^{15}$. Such ensemble size and cooperativity are achievable based on prior experimental works \cite{eisenach2021cavity,angerer2017ultralong}. The sensitivity is independent of the $g_s^2/\kappa$ ratio for fixed cooperativity, and further sensitivity dependence on spin-cavity parameters is detailed in Supplementary Information \cite{Supplementary}.

\textit{Comagnetometry - }One of the main challenges for spin-based gyroscopes in ambient conditions is the magnetic field environment. The magnetic field generates a varying nuclear phase via $\gamma_n B_z$, indistinguishable from the rotation signal in the Hamiltonian, therefore setting the noise floor $\eta_r$. The concept of a comagnetometer involves utilizing the electron spin to measure the magnetic field, which is then used to compensate for nuclear spin phase accumulation induced by the magnetic environment \cite{soshenko2021nuclear}.

Here, we apply a second probe tone to measure the electron spin frequency similarly to previous cQED work \cite{wang2024spin}. Fig. 3(a) shows cavity reflection $|r|^2$ as a function of $\Delta$ and $\Delta_s$. Three avoid-crossing features are observed, corresponding to the three $^{14}N$ hyperfine transitions assuming each subensemble has an equal population with $C = 10$. We plot the magnetic sensitivity-limited ARW floor produced by the second probe tone as a function of its detuning $\Delta$ in Fig. 3(a), bottom. With the two-field interference applied in the $|1\rangle$, $|2\rangle$ subspace, however, the coupling strengths become imbalanced. The driving field polarizes the nuclear spin transition to state $|1\rangle$ from $|2\rangle$, reducing the coupling strength for the $2_+$ transition compared to $1_+$ and producing an asymmetry in the sensitivity spectrum. The comagnetometer sensitivity maximized at a probe detuning of $\Delta = -4.7$ MHz (point I in Fig. 3(a)), where the noise floor scales with comagnetometer probe power as $\eta_r$ = 1 mdeg/s/$\mathrm{\sqrt{Hz}}$ with power $P_2 = -40$ dBm. This resonates with the $3_+$ transition, effectively using the third nuclear spin hyperfine subensemble ($m_I = +1$) as a comagnetometer.

The sensitivity degradation induced by the second probe, detuned from the EIT probe by a frequency detuning $\Delta_T$ and Rabi frequency $\Omega_R$, is plotted in Fig. 3(b) \cite{Supplementary}. The ideal co-magnetometry operating point is shown by the red dot I, with a Rabi frequency $\Omega_R = 4.4$ kHz producing a $\eta_r$ = 1 mdeg/s/$\mathrm{\sqrt{Hz}}$. At this point, the EIT rotational sensitivity is not affected by the comagnetometer probe, with $\Delta\eta/\eta < 1\%$. The compensation for the magnetic field drift also provides a direct drift correction, providing potential for long-term stability for the nNV-cQED system.

\textit{Vector gyroscope - }We extend the nNV-cQED system to a vector gyroscope with a multifield EIT scheme. The external magnetic bias is adjusted such that the electron spin transitions are near the cavity resonance yet resolvable ($E_i - E_j > \Gamma$), with multiple fields applied for each spin subensemble. Each subensemble, with the same cooperativity as in the previous section, measures the rotation projection on its axis, as the external bias is small compared to the electron zero-field splitting. Subensembles must be densely packed within the cavity linewidth since the rotation signal $S$ drops significantly beyond, as shown in Fig. 3(c).

Two kinds of crosstalk may affect the nNV-cQED vector device. First, since the EIT condition is simultaneously met for all subensembles detuned in common-mode frequency, we need to ensure that the signal measured from a particular EIT probe can be associated with shifts of one ensemble. To this end, we consider the signal response $S$ from an EIT probe at frequency $\Delta_\mathrm{C}$ associated with an energy shift of each subensemble, given that the ensembles are separated by 0.8 MHz. As shown in Fig. 3(d), an EIT probe on resonance with the central ensemble under these conditions would have a maximum relative sensitivity of 10\% to the nearest off-resonant ensembles (typically 0.5\%). A linear crosstalk elimination process can be performed for further correction.

Second, the comagnetometer and multiple closely-packed EIT probes will induce additional interference effects that degrade performance, as discussed in the comagnetometer session. The sensitivity degradation is also within 1\% indicated in Fig. 3(b) (dots labeled $E_i$).

\textit{Conclusion and Discussion - } Our proposed nNV-cQED gyroscope scheme shows that nuclear spin readout can performed using a strongly coupled microwave cavity, with readout fidelity of $\sigma_n = 37$ outperforming current ensemble optical approaches. The extension of nNV-cQED to multiple ensembles enables co-magnetometry and vector sensing, which is not currently achieved in atomic inertial sensors or NV electron-spin cQED experiments. Finally, we show that the nNV-cQED system operating with high cooperativity enters the counterintuitive MWI regime, potentially into an oscillatory phase. 

Future work will explore extending the two-field theory to inhomogeneous ensembles, especially in nonlinear, bistable, and MWI regimes; optimizing spin refrigeration and system parameters to approach the quantum limit; and alternative modalities, including applications in microwave memories and timekeeping \cite{wei2020broadband,trusheim2020polariton}.


\textit{Acknowledgment - }The authors would like to thank Bo-Han Wu for fruitful discussions. H.W.~acknowledges support from Bosch Inc. S.W. acknowledges support from Honda Research Institute USA, Inc. Y.D. acknowledges support from Mathworks Fellowship. D.R.E.~acknowledges funding from the MITRE Corporation and the U.S.~NSF Center for Ultracold Atoms.

\bibliography{apssamp}

\providecommand{\noopsort}[1]{}\providecommand{\singleletter}[1]{#1}%
\begin{thebibliography}{38}%
\makeatletter
\providecommand \@ifxundefined [1]{%
 \@ifx{#1\undefined}
}%
\providecommand \@ifnum [1]{%
 \ifnum #1\expandafter \@firstoftwo
 \else \expandafter \@secondoftwo
 \fi
}%
\providecommand \@ifx [1]{%
 \ifx #1\expandafter \@firstoftwo
 \else \expandafter \@secondoftwo
 \fi
}%
\providecommand \natexlab [1]{#1}%
\providecommand \enquote  [1]{``#1''}%
\providecommand \bibnamefont  [1]{#1}%
\providecommand \bibfnamefont [1]{#1}%
\providecommand \citenamefont [1]{#1}%
\providecommand \href@noop [0]{\@secondoftwo}%
\providecommand \href [0]{\begingroup \@sanitize@url \@href}%
\providecommand \@href[1]{\@@startlink{#1}\@@href}%
\providecommand \@@href[1]{\endgroup#1\@@endlink}%
\providecommand \@sanitize@url [0]{\catcode `\\12\catcode `\$12\catcode `\&12\catcode `\#12\catcode `\^12\catcode `\_12\catcode `\%12\relax}%
\providecommand \@@startlink[1]{}%
\providecommand \@@endlink[0]{}%
\providecommand \url  [0]{\begingroup\@sanitize@url \@url }%
\providecommand \@url [1]{\endgroup\@href {#1}{\urlprefix }}%
\providecommand \urlprefix  [0]{URL }%
\providecommand \Eprint [0]{\href }%
\providecommand \doibase [0]{https://doi.org/}%
\providecommand \selectlanguage [0]{\@gobble}%
\providecommand \bibinfo  [0]{\@secondoftwo}%
\providecommand \bibfield  [0]{\@secondoftwo}%
\providecommand \translation [1]{[#1]}%
\providecommand \BibitemOpen [0]{}%
\providecommand \bibitemStop [0]{}%
\providecommand \bibitemNoStop [0]{.\EOS\space}%
\providecommand \EOS [0]{\spacefactor3000\relax}%
\providecommand \BibitemShut  [1]{\csname bibitem#1\endcsname}%
\let\auto@bib@innerbib\@empty
\bibitem [{\citenamefont {Barry}\ \emph {et~al.}(2020)\citenamefont {Barry}, \citenamefont {Schloss}, \citenamefont {Bauch}, \citenamefont {Turner}, \citenamefont {Hart}, \citenamefont {Pham},\ and\ \citenamefont {Walsworth}}]{barry2020sensitivity}%
  \BibitemOpen
  \bibfield  {author} {\bibinfo {author} {\bibfnamefont {J.~F.}\ \bibnamefont {Barry}}, \bibinfo {author} {\bibfnamefont {J.~M.}\ \bibnamefont {Schloss}}, \bibinfo {author} {\bibfnamefont {E.}~\bibnamefont {Bauch}}, \bibinfo {author} {\bibfnamefont {M.~J.}\ \bibnamefont {Turner}}, \bibinfo {author} {\bibfnamefont {C.~A.}\ \bibnamefont {Hart}}, \bibinfo {author} {\bibfnamefont {L.~M.}\ \bibnamefont {Pham}},\ and\ \bibinfo {author} {\bibfnamefont {R.~L.}\ \bibnamefont {Walsworth}},\ }\bibfield  {title} {\bibinfo {title} {Sensitivity optimization for nv-diamond magnetometry},\ }\href@noop {} {\bibfield  {journal} {\bibinfo  {journal} {Reviews of Modern Physics}\ }\textbf {\bibinfo {volume} {92}},\ \bibinfo {pages} {015004} (\bibinfo {year} {2020})}\BibitemShut {NoStop}%
\bibitem [{\citenamefont {Degen}\ \emph {et~al.}(2017)\citenamefont {Degen}, \citenamefont {Reinhard},\ and\ \citenamefont {Cappellaro}}]{degen2017quantum}%
  \BibitemOpen
  \bibfield  {author} {\bibinfo {author} {\bibfnamefont {C.~L.}\ \bibnamefont {Degen}}, \bibinfo {author} {\bibfnamefont {F.}~\bibnamefont {Reinhard}},\ and\ \bibinfo {author} {\bibfnamefont {P.}~\bibnamefont {Cappellaro}},\ }\bibfield  {title} {\bibinfo {title} {Quantum sensing},\ }\href@noop {} {\bibfield  {journal} {\bibinfo  {journal} {Reviews of modern physics}\ }\textbf {\bibinfo {volume} {89}},\ \bibinfo {pages} {035002} (\bibinfo {year} {2017})}\BibitemShut {NoStop}%
\bibitem [{\citenamefont {Doherty}\ \emph {et~al.}(2013)\citenamefont {Doherty}, \citenamefont {Manson}, \citenamefont {Delaney}, \citenamefont {Jelezko}, \citenamefont {Wrachtrup},\ and\ \citenamefont {Hollenberg}}]{doherty2013nitrogen}%
  \BibitemOpen
  \bibfield  {author} {\bibinfo {author} {\bibfnamefont {M.~W.}\ \bibnamefont {Doherty}}, \bibinfo {author} {\bibfnamefont {N.~B.}\ \bibnamefont {Manson}}, \bibinfo {author} {\bibfnamefont {P.}~\bibnamefont {Delaney}}, \bibinfo {author} {\bibfnamefont {F.}~\bibnamefont {Jelezko}}, \bibinfo {author} {\bibfnamefont {J.}~\bibnamefont {Wrachtrup}},\ and\ \bibinfo {author} {\bibfnamefont {L.~C.}\ \bibnamefont {Hollenberg}},\ }\bibfield  {title} {\bibinfo {title} {The nitrogen-vacancy colour centre in diamond},\ }\href@noop {} {\bibfield  {journal} {\bibinfo  {journal} {Physics Reports}\ }\textbf {\bibinfo {volume} {528}},\ \bibinfo {pages} {1} (\bibinfo {year} {2013})}\BibitemShut {NoStop}%
\bibitem [{\citenamefont {Kim}\ \emph {et~al.}(2023)\citenamefont {Kim}, \citenamefont {Choi}, \citenamefont {Trusheim}, \citenamefont {Wang},\ and\ \citenamefont {Englund}}]{kim2023nanophotonic}%
  \BibitemOpen
  \bibfield  {author} {\bibinfo {author} {\bibfnamefont {L.}~\bibnamefont {Kim}}, \bibinfo {author} {\bibfnamefont {H.}~\bibnamefont {Choi}}, \bibinfo {author} {\bibfnamefont {M.~E.}\ \bibnamefont {Trusheim}}, \bibinfo {author} {\bibfnamefont {H.}~\bibnamefont {Wang}},\ and\ \bibinfo {author} {\bibfnamefont {D.~R.}\ \bibnamefont {Englund}},\ }\bibfield  {title} {\bibinfo {title} {Nanophotonic quantum sensing with engineered spin-optic coupling},\ }\href@noop {} {\bibfield  {journal} {\bibinfo  {journal} {Nanophotonics}\ } (\bibinfo {year} {2023})}\BibitemShut {NoStop}%
\bibitem [{\citenamefont {Wang}\ \emph {et~al.}(2023)\citenamefont {Wang}, \citenamefont {Trusheim}, \citenamefont {Kim}, \citenamefont {Raniwala},\ and\ \citenamefont {Englund}}]{wang2023field}%
  \BibitemOpen
  \bibfield  {author} {\bibinfo {author} {\bibfnamefont {H.}~\bibnamefont {Wang}}, \bibinfo {author} {\bibfnamefont {M.~E.}\ \bibnamefont {Trusheim}}, \bibinfo {author} {\bibfnamefont {L.}~\bibnamefont {Kim}}, \bibinfo {author} {\bibfnamefont {H.}~\bibnamefont {Raniwala}},\ and\ \bibinfo {author} {\bibfnamefont {D.~R.}\ \bibnamefont {Englund}},\ }\bibfield  {title} {\bibinfo {title} {Field programmable spin arrays for scalable quantum repeaters},\ }\href@noop {} {\bibfield  {journal} {\bibinfo  {journal} {Nature Communications}\ }\textbf {\bibinfo {volume} {14}},\ \bibinfo {pages} {704} (\bibinfo {year} {2023})}\BibitemShut {NoStop}%
\bibitem [{\citenamefont {Du}\ \emph {et~al.}(2017)\citenamefont {Du}, \citenamefont {Van~der Sar}, \citenamefont {Zhou}, \citenamefont {Upadhyaya}, \citenamefont {Casola}, \citenamefont {Zhang}, \citenamefont {Onbasli}, \citenamefont {Ross}, \citenamefont {Walsworth}, \citenamefont {Tserkovnyak} \emph {et~al.}}]{du2017control}%
  \BibitemOpen
  \bibfield  {author} {\bibinfo {author} {\bibfnamefont {C.}~\bibnamefont {Du}}, \bibinfo {author} {\bibfnamefont {T.}~\bibnamefont {Van~der Sar}}, \bibinfo {author} {\bibfnamefont {T.~X.}\ \bibnamefont {Zhou}}, \bibinfo {author} {\bibfnamefont {P.}~\bibnamefont {Upadhyaya}}, \bibinfo {author} {\bibfnamefont {F.}~\bibnamefont {Casola}}, \bibinfo {author} {\bibfnamefont {H.}~\bibnamefont {Zhang}}, \bibinfo {author} {\bibfnamefont {M.~C.}\ \bibnamefont {Onbasli}}, \bibinfo {author} {\bibfnamefont {C.~A.}\ \bibnamefont {Ross}}, \bibinfo {author} {\bibfnamefont {R.~L.}\ \bibnamefont {Walsworth}}, \bibinfo {author} {\bibfnamefont {Y.}~\bibnamefont {Tserkovnyak}}, \emph {et~al.},\ }\bibfield  {title} {\bibinfo {title} {Control and local measurement of the spin chemical potential in a magnetic insulator},\ }\href@noop {} {\bibfield  {journal} {\bibinfo  {journal} {Science}\ }\textbf {\bibinfo {volume} {357}},\ \bibinfo {pages} {195} (\bibinfo {year} {2017})}\BibitemShut {NoStop}%
\bibitem [{\citenamefont {Pelliccione}\ \emph {et~al.}(2016)\citenamefont {Pelliccione}, \citenamefont {Jenkins}, \citenamefont {Ovartchaiyapong}, \citenamefont {Reetz}, \citenamefont {Emmanouilidou}, \citenamefont {Ni},\ and\ \citenamefont {Bleszynski~Jayich}}]{pelliccione2016scanned}%
  \BibitemOpen
  \bibfield  {author} {\bibinfo {author} {\bibfnamefont {M.}~\bibnamefont {Pelliccione}}, \bibinfo {author} {\bibfnamefont {A.}~\bibnamefont {Jenkins}}, \bibinfo {author} {\bibfnamefont {P.}~\bibnamefont {Ovartchaiyapong}}, \bibinfo {author} {\bibfnamefont {C.}~\bibnamefont {Reetz}}, \bibinfo {author} {\bibfnamefont {E.}~\bibnamefont {Emmanouilidou}}, \bibinfo {author} {\bibfnamefont {N.}~\bibnamefont {Ni}},\ and\ \bibinfo {author} {\bibfnamefont {A.~C.}\ \bibnamefont {Bleszynski~Jayich}},\ }\bibfield  {title} {\bibinfo {title} {Scanned probe imaging of nanoscale magnetism at cryogenic temperatures with a single-spin quantum sensor},\ }\href@noop {} {\bibfield  {journal} {\bibinfo  {journal} {Nature nanotechnology}\ }\textbf {\bibinfo {volume} {11}},\ \bibinfo {pages} {700} (\bibinfo {year} {2016})}\BibitemShut {NoStop}%
\bibitem [{\citenamefont {Hu}\ \emph {et~al.}(2024)\citenamefont {Hu}, \citenamefont {Nagle}, \citenamefont {Duan}, \citenamefont {Wang},\ and\ \citenamefont {Englund}}]{hu2024developing}%
  \BibitemOpen
  \bibfield  {author} {\bibinfo {author} {\bibfnamefont {Y.}~\bibnamefont {Hu}}, \bibinfo {author} {\bibfnamefont {S.}~\bibnamefont {Nagle}}, \bibinfo {author} {\bibfnamefont {Y.}~\bibnamefont {Duan}}, \bibinfo {author} {\bibfnamefont {H.}~\bibnamefont {Wang}},\ and\ \bibinfo {author} {\bibfnamefont {D.}~\bibnamefont {Englund}},\ }\href@noop {} {\bibinfo {title} {Developing 3d models of atom-like defect spin memories in crystals for quantum technology research and education}} (\bibinfo {year} {2024})\BibitemShut {NoStop}%
\bibitem [{\citenamefont {Bauch}\ \emph {et~al.}(2018)\citenamefont {Bauch}, \citenamefont {Hart}, \citenamefont {Schloss}, \citenamefont {Turner}, \citenamefont {Barry}, \citenamefont {Kehayias}, \citenamefont {Singh},\ and\ \citenamefont {Walsworth}}]{bauch2018ultralong}%
  \BibitemOpen
  \bibfield  {author} {\bibinfo {author} {\bibfnamefont {E.}~\bibnamefont {Bauch}}, \bibinfo {author} {\bibfnamefont {C.~A.}\ \bibnamefont {Hart}}, \bibinfo {author} {\bibfnamefont {J.~M.}\ \bibnamefont {Schloss}}, \bibinfo {author} {\bibfnamefont {M.~J.}\ \bibnamefont {Turner}}, \bibinfo {author} {\bibfnamefont {J.~F.}\ \bibnamefont {Barry}}, \bibinfo {author} {\bibfnamefont {P.}~\bibnamefont {Kehayias}}, \bibinfo {author} {\bibfnamefont {S.}~\bibnamefont {Singh}},\ and\ \bibinfo {author} {\bibfnamefont {R.~L.}\ \bibnamefont {Walsworth}},\ }\bibfield  {title} {\bibinfo {title} {Ultralong dephasing times in solid-state spin ensembles via quantum control},\ }\href@noop {} {\bibfield  {journal} {\bibinfo  {journal} {Physical Review X}\ }\textbf {\bibinfo {volume} {8}},\ \bibinfo {pages} {031025} (\bibinfo {year} {2018})}\BibitemShut {NoStop}%
\bibitem [{\citenamefont {Stanwix}\ \emph {et~al.}(2010)\citenamefont {Stanwix}, \citenamefont {Pham}, \citenamefont {Maze}, \citenamefont {Le~Sage}, \citenamefont {Yeung}, \citenamefont {Cappellaro}, \citenamefont {Hemmer}, \citenamefont {Yacoby}, \citenamefont {Lukin},\ and\ \citenamefont {Walsworth}}]{stanwix2010coherence}%
  \BibitemOpen
  \bibfield  {author} {\bibinfo {author} {\bibfnamefont {P.~L.}\ \bibnamefont {Stanwix}}, \bibinfo {author} {\bibfnamefont {L.~M.}\ \bibnamefont {Pham}}, \bibinfo {author} {\bibfnamefont {J.~R.}\ \bibnamefont {Maze}}, \bibinfo {author} {\bibfnamefont {D.}~\bibnamefont {Le~Sage}}, \bibinfo {author} {\bibfnamefont {T.~K.}\ \bibnamefont {Yeung}}, \bibinfo {author} {\bibfnamefont {P.}~\bibnamefont {Cappellaro}}, \bibinfo {author} {\bibfnamefont {P.~R.}\ \bibnamefont {Hemmer}}, \bibinfo {author} {\bibfnamefont {A.}~\bibnamefont {Yacoby}}, \bibinfo {author} {\bibfnamefont {M.~D.}\ \bibnamefont {Lukin}},\ and\ \bibinfo {author} {\bibfnamefont {R.~L.}\ \bibnamefont {Walsworth}},\ }\bibfield  {title} {\bibinfo {title} {Coherence of nitrogen-vacancy electronic spin ensembles in diamond},\ }\href@noop {} {\bibfield  {journal} {\bibinfo  {journal} {Physical Review B}\ }\textbf {\bibinfo {volume} {82}},\ \bibinfo {pages} {201201} (\bibinfo {year} {2010})}\BibitemShut {NoStop}%
\bibitem [{\citenamefont {Wang}\ \emph {et~al.}(2021)\citenamefont {Wang}, \citenamefont {Li},\ and\ \citenamefont {Cappellaro}}]{wang2021observation}%
  \BibitemOpen
  \bibfield  {author} {\bibinfo {author} {\bibfnamefont {G.}~\bibnamefont {Wang}}, \bibinfo {author} {\bibfnamefont {C.}~\bibnamefont {Li}},\ and\ \bibinfo {author} {\bibfnamefont {P.}~\bibnamefont {Cappellaro}},\ }\bibfield  {title} {\bibinfo {title} {Observation of symmetry-protected selection rules in periodically driven quantum systems},\ }\href@noop {} {\bibfield  {journal} {\bibinfo  {journal} {Physical Review Letters}\ }\textbf {\bibinfo {volume} {127}},\ \bibinfo {pages} {140604} (\bibinfo {year} {2021})}\BibitemShut {NoStop}%
\bibitem [{\citenamefont {Kalb}\ \emph {et~al.}(2018)\citenamefont {Kalb}, \citenamefont {Humphreys}, \citenamefont {Slim},\ and\ \citenamefont {Hanson}}]{kalb2018dephasing}%
  \BibitemOpen
  \bibfield  {author} {\bibinfo {author} {\bibfnamefont {N.}~\bibnamefont {Kalb}}, \bibinfo {author} {\bibfnamefont {P.~C.}\ \bibnamefont {Humphreys}}, \bibinfo {author} {\bibfnamefont {J.}~\bibnamefont {Slim}},\ and\ \bibinfo {author} {\bibfnamefont {R.}~\bibnamefont {Hanson}},\ }\bibfield  {title} {\bibinfo {title} {Dephasing mechanisms of diamond-based nuclear-spin memories for quantum networks},\ }\href@noop {} {\bibfield  {journal} {\bibinfo  {journal} {Physical Review A}\ }\textbf {\bibinfo {volume} {97}},\ \bibinfo {pages} {062330} (\bibinfo {year} {2018})}\BibitemShut {NoStop}%
\bibitem [{\citenamefont {Jarmola}\ \emph {et~al.}(2021)\citenamefont {Jarmola}, \citenamefont {Lourette}, \citenamefont {Acosta}, \citenamefont {Birdwell}, \citenamefont {Bl{\"u}mler}, \citenamefont {Budker}, \citenamefont {Ivanov},\ and\ \citenamefont {Malinovsky}}]{jarmola2021demonstration}%
  \BibitemOpen
  \bibfield  {author} {\bibinfo {author} {\bibfnamefont {A.}~\bibnamefont {Jarmola}}, \bibinfo {author} {\bibfnamefont {S.}~\bibnamefont {Lourette}}, \bibinfo {author} {\bibfnamefont {V.~M.}\ \bibnamefont {Acosta}}, \bibinfo {author} {\bibfnamefont {A.~G.}\ \bibnamefont {Birdwell}}, \bibinfo {author} {\bibfnamefont {P.}~\bibnamefont {Bl{\"u}mler}}, \bibinfo {author} {\bibfnamefont {D.}~\bibnamefont {Budker}}, \bibinfo {author} {\bibfnamefont {T.}~\bibnamefont {Ivanov}},\ and\ \bibinfo {author} {\bibfnamefont {V.~S.}\ \bibnamefont {Malinovsky}},\ }\bibfield  {title} {\bibinfo {title} {Demonstration of diamond nuclear spin gyroscope},\ }\href@noop {} {\bibfield  {journal} {\bibinfo  {journal} {Science advances}\ }\textbf {\bibinfo {volume} {7}},\ \bibinfo {pages} {eabl3840} (\bibinfo {year} {2021})}\BibitemShut {NoStop}%
\bibitem [{\citenamefont {Wang}\ \emph {et~al.}(2024{\natexlab{a}})\citenamefont {Wang}, \citenamefont {Nguyen},\ and\ \citenamefont {Cappellaro}}]{wang2024hyperfine}%
  \BibitemOpen
  \bibfield  {author} {\bibinfo {author} {\bibfnamefont {G.}~\bibnamefont {Wang}}, \bibinfo {author} {\bibfnamefont {M.-T.}\ \bibnamefont {Nguyen}},\ and\ \bibinfo {author} {\bibfnamefont {P.}~\bibnamefont {Cappellaro}},\ }\bibfield  {title} {\bibinfo {title} {Hyperfine-enhanced gyroscope based on solid-state spins},\ }\href@noop {} {\bibfield  {journal} {\bibinfo  {journal} {Physical Review Letters}\ }\textbf {\bibinfo {volume} {133}},\ \bibinfo {pages} {150801} (\bibinfo {year} {2024}{\natexlab{a}})}\BibitemShut {NoStop}%
\bibitem [{\citenamefont {Wang}\ \emph {et~al.}(2024{\natexlab{b}})\citenamefont {Wang}, \citenamefont {Nguyen}, \citenamefont {deQuilettes}, \citenamefont {Price}, \citenamefont {Hu}, \citenamefont {Braje},\ and\ \citenamefont {Cappellaro}}]{wang2024emulated}%
  \BibitemOpen
  \bibfield  {author} {\bibinfo {author} {\bibfnamefont {G.}~\bibnamefont {Wang}}, \bibinfo {author} {\bibfnamefont {M.-T.}\ \bibnamefont {Nguyen}}, \bibinfo {author} {\bibfnamefont {D.~W.}\ \bibnamefont {deQuilettes}}, \bibinfo {author} {\bibfnamefont {E.}~\bibnamefont {Price}}, \bibinfo {author} {\bibfnamefont {Z.}~\bibnamefont {Hu}}, \bibinfo {author} {\bibfnamefont {D.~A.}\ \bibnamefont {Braje}},\ and\ \bibinfo {author} {\bibfnamefont {P.}~\bibnamefont {Cappellaro}},\ }\bibfield  {title} {\bibinfo {title} {Emulated nuclear spin gyroscope with 15 n-v centers in diamond},\ }\href@noop {} {\bibfield  {journal} {\bibinfo  {journal} {Physical Review Applied}\ }\textbf {\bibinfo {volume} {22}},\ \bibinfo {pages} {044016} (\bibinfo {year} {2024}{\natexlab{b}})}\BibitemShut {NoStop}%
\bibitem [{\citenamefont {Stockton}\ \emph {et~al.}(2011)\citenamefont {Stockton}, \citenamefont {Takase},\ and\ \citenamefont {Kasevich}}]{stockton2011absolute}%
  \BibitemOpen
  \bibfield  {author} {\bibinfo {author} {\bibfnamefont {J.}~\bibnamefont {Stockton}}, \bibinfo {author} {\bibfnamefont {K.}~\bibnamefont {Takase}},\ and\ \bibinfo {author} {\bibfnamefont {M.}~\bibnamefont {Kasevich}},\ }\bibfield  {title} {\bibinfo {title} {Absolute geodetic rotation measurement using atom interferometry},\ }\href@noop {} {\bibfield  {journal} {\bibinfo  {journal} {Physical review letters}\ }\textbf {\bibinfo {volume} {107}},\ \bibinfo {pages} {133001} (\bibinfo {year} {2011})}\BibitemShut {NoStop}%
\bibitem [{\citenamefont {Ledbetter}\ \emph {et~al.}(2012)\citenamefont {Ledbetter}, \citenamefont {Jensen}, \citenamefont {Fischer}, \citenamefont {Jarmola},\ and\ \citenamefont {Budker}}]{ledbetter2012gyroscopes}%
  \BibitemOpen
  \bibfield  {author} {\bibinfo {author} {\bibfnamefont {M.}~\bibnamefont {Ledbetter}}, \bibinfo {author} {\bibfnamefont {K.}~\bibnamefont {Jensen}}, \bibinfo {author} {\bibfnamefont {R.}~\bibnamefont {Fischer}}, \bibinfo {author} {\bibfnamefont {A.}~\bibnamefont {Jarmola}},\ and\ \bibinfo {author} {\bibfnamefont {D.}~\bibnamefont {Budker}},\ }\bibfield  {title} {\bibinfo {title} {Gyroscopes based on nitrogen-vacancy centers in diamond},\ }\href@noop {} {\bibfield  {journal} {\bibinfo  {journal} {Physical Review A}\ }\textbf {\bibinfo {volume} {86}},\ \bibinfo {pages} {052116} (\bibinfo {year} {2012})}\BibitemShut {NoStop}%
\bibitem [{\citenamefont {Maclaurin}\ \emph {et~al.}(2012)\citenamefont {Maclaurin}, \citenamefont {Doherty}, \citenamefont {Hollenberg},\ and\ \citenamefont {Martin}}]{maclaurin2012measurable}%
  \BibitemOpen
  \bibfield  {author} {\bibinfo {author} {\bibfnamefont {D.}~\bibnamefont {Maclaurin}}, \bibinfo {author} {\bibfnamefont {M.}~\bibnamefont {Doherty}}, \bibinfo {author} {\bibfnamefont {L.}~\bibnamefont {Hollenberg}},\ and\ \bibinfo {author} {\bibfnamefont {A.}~\bibnamefont {Martin}},\ }\bibfield  {title} {\bibinfo {title} {Measurable quantum geometric phase from a rotating single spin},\ }\href@noop {} {\bibfield  {journal} {\bibinfo  {journal} {Physical review letters}\ }\textbf {\bibinfo {volume} {108}},\ \bibinfo {pages} {240403} (\bibinfo {year} {2012})}\BibitemShut {NoStop}%
\bibitem [{\citenamefont {Ajoy}\ and\ \citenamefont {Cappellaro}(2012)}]{ajoy2012stable}%
  \BibitemOpen
  \bibfield  {author} {\bibinfo {author} {\bibfnamefont {A.}~\bibnamefont {Ajoy}}\ and\ \bibinfo {author} {\bibfnamefont {P.}~\bibnamefont {Cappellaro}},\ }\bibfield  {title} {\bibinfo {title} {Stable three-axis nuclear-spin gyroscope in diamond},\ }\href@noop {} {\bibfield  {journal} {\bibinfo  {journal} {Physical Review A}\ }\textbf {\bibinfo {volume} {86}},\ \bibinfo {pages} {062104} (\bibinfo {year} {2012})}\BibitemShut {NoStop}%
\bibitem [{\citenamefont {Lai}\ \emph {et~al.}(2020)\citenamefont {Lai}, \citenamefont {Suh}, \citenamefont {Lu}, \citenamefont {Shen}, \citenamefont {Yang}, \citenamefont {Wang}, \citenamefont {Li}, \citenamefont {Lee}, \citenamefont {Yang},\ and\ \citenamefont {Vahala}}]{lai2020earth}%
  \BibitemOpen
  \bibfield  {author} {\bibinfo {author} {\bibfnamefont {Y.-H.}\ \bibnamefont {Lai}}, \bibinfo {author} {\bibfnamefont {M.-G.}\ \bibnamefont {Suh}}, \bibinfo {author} {\bibfnamefont {Y.-K.}\ \bibnamefont {Lu}}, \bibinfo {author} {\bibfnamefont {B.}~\bibnamefont {Shen}}, \bibinfo {author} {\bibfnamefont {Q.-F.}\ \bibnamefont {Yang}}, \bibinfo {author} {\bibfnamefont {H.}~\bibnamefont {Wang}}, \bibinfo {author} {\bibfnamefont {J.}~\bibnamefont {Li}}, \bibinfo {author} {\bibfnamefont {S.~H.}\ \bibnamefont {Lee}}, \bibinfo {author} {\bibfnamefont {K.~Y.}\ \bibnamefont {Yang}},\ and\ \bibinfo {author} {\bibfnamefont {K.}~\bibnamefont {Vahala}},\ }\bibfield  {title} {\bibinfo {title} {Earth rotation measured by a chip-scale ring laser gyroscope},\ }\href@noop {} {\bibfield  {journal} {\bibinfo  {journal} {Nature Photonics}\ }\textbf {\bibinfo {volume} {14}},\ \bibinfo {pages} {345} (\bibinfo {year} {2020})}\BibitemShut {NoStop}%
\bibitem [{\citenamefont {Hokmabadi}\ \emph {et~al.}(2019)\citenamefont {Hokmabadi}, \citenamefont {Schumer}, \citenamefont {Christodoulides},\ and\ \citenamefont {Khajavikhan}}]{hokmabadi2019non}%
  \BibitemOpen
  \bibfield  {author} {\bibinfo {author} {\bibfnamefont {M.~P.}\ \bibnamefont {Hokmabadi}}, \bibinfo {author} {\bibfnamefont {A.}~\bibnamefont {Schumer}}, \bibinfo {author} {\bibfnamefont {D.~N.}\ \bibnamefont {Christodoulides}},\ and\ \bibinfo {author} {\bibfnamefont {M.}~\bibnamefont {Khajavikhan}},\ }\bibfield  {title} {\bibinfo {title} {Non-hermitian ring laser gyroscopes with enhanced sagnac sensitivity},\ }\href@noop {} {\bibfield  {journal} {\bibinfo  {journal} {Nature}\ }\textbf {\bibinfo {volume} {576}},\ \bibinfo {pages} {70} (\bibinfo {year} {2019})}\BibitemShut {NoStop}%
\bibitem [{\citenamefont {Soshenko}\ \emph {et~al.}(2021)\citenamefont {Soshenko}, \citenamefont {Bolshedvorskii}, \citenamefont {Rubinas}, \citenamefont {Sorokin}, \citenamefont {Smolyaninov}, \citenamefont {Vorobyov},\ and\ \citenamefont {Akimov}}]{soshenko2021nuclear}%
  \BibitemOpen
  \bibfield  {author} {\bibinfo {author} {\bibfnamefont {V.~V.}\ \bibnamefont {Soshenko}}, \bibinfo {author} {\bibfnamefont {S.~V.}\ \bibnamefont {Bolshedvorskii}}, \bibinfo {author} {\bibfnamefont {O.}~\bibnamefont {Rubinas}}, \bibinfo {author} {\bibfnamefont {V.~N.}\ \bibnamefont {Sorokin}}, \bibinfo {author} {\bibfnamefont {A.~N.}\ \bibnamefont {Smolyaninov}}, \bibinfo {author} {\bibfnamefont {V.~V.}\ \bibnamefont {Vorobyov}},\ and\ \bibinfo {author} {\bibfnamefont {A.~V.}\ \bibnamefont {Akimov}},\ }\bibfield  {title} {\bibinfo {title} {Nuclear spin gyroscope based on the nitrogen vacancy center in diamond},\ }\href@noop {} {\bibfield  {journal} {\bibinfo  {journal} {Physical Review Letters}\ }\textbf {\bibinfo {volume} {126}},\ \bibinfo {pages} {197702} (\bibinfo {year} {2021})}\BibitemShut {NoStop}%
\bibitem [{\citenamefont {Wang}\ \emph {et~al.}(2024{\natexlab{c}})\citenamefont {Wang}, \citenamefont {Tiwari}, \citenamefont {Jacobs}, \citenamefont {Judy}, \citenamefont {Zhang}, \citenamefont {Englund},\ and\ \citenamefont {Trusheim}}]{wang2024spin}%
  \BibitemOpen
  \bibfield  {author} {\bibinfo {author} {\bibfnamefont {H.}~\bibnamefont {Wang}}, \bibinfo {author} {\bibfnamefont {K.~L.}\ \bibnamefont {Tiwari}}, \bibinfo {author} {\bibfnamefont {K.}~\bibnamefont {Jacobs}}, \bibinfo {author} {\bibfnamefont {M.}~\bibnamefont {Judy}}, \bibinfo {author} {\bibfnamefont {X.}~\bibnamefont {Zhang}}, \bibinfo {author} {\bibfnamefont {D.~R.}\ \bibnamefont {Englund}},\ and\ \bibinfo {author} {\bibfnamefont {M.~E.}\ \bibnamefont {Trusheim}},\ }\bibfield  {title} {\bibinfo {title} {A spin-refrigerated cavity quantum electrodynamic sensor},\ }\href@noop {} {\bibfield  {journal} {\bibinfo  {journal} {Nature Communications}\ }\textbf {\bibinfo {volume} {15}},\ \bibinfo {pages} {10320} (\bibinfo {year} {2024}{\natexlab{c}})}\BibitemShut {NoStop}%
\bibitem [{\citenamefont {Eisenach}\ \emph {et~al.}(2021)\citenamefont {Eisenach}, \citenamefont {Barry}, \citenamefont {O’Keeffe}, \citenamefont {Schloss}, \citenamefont {Steinecker}, \citenamefont {Englund},\ and\ \citenamefont {Braje}}]{eisenach2021cavity}%
  \BibitemOpen
  \bibfield  {author} {\bibinfo {author} {\bibfnamefont {E.~R.}\ \bibnamefont {Eisenach}}, \bibinfo {author} {\bibfnamefont {J.~F.}\ \bibnamefont {Barry}}, \bibinfo {author} {\bibfnamefont {M.~F.}\ \bibnamefont {O’Keeffe}}, \bibinfo {author} {\bibfnamefont {J.~M.}\ \bibnamefont {Schloss}}, \bibinfo {author} {\bibfnamefont {M.~H.}\ \bibnamefont {Steinecker}}, \bibinfo {author} {\bibfnamefont {D.~R.}\ \bibnamefont {Englund}},\ and\ \bibinfo {author} {\bibfnamefont {D.~A.}\ \bibnamefont {Braje}},\ }\bibfield  {title} {\bibinfo {title} {Cavity-enhanced microwave readout of a solid-state spin sensor},\ }\href@noop {} {\bibfield  {journal} {\bibinfo  {journal} {Nature communications}\ }\textbf {\bibinfo {volume} {12}},\ \bibinfo {pages} {1357} (\bibinfo {year} {2021})}\BibitemShut {NoStop}%
\bibitem [{\citenamefont {Zhu}(1992)}]{zhu1992lasing}%
  \BibitemOpen
  \bibfield  {author} {\bibinfo {author} {\bibfnamefont {Y.}~\bibnamefont {Zhu}},\ }\bibfield  {title} {\bibinfo {title} {Lasing without inversion in a closed three-level system},\ }\href@noop {} {\bibfield  {journal} {\bibinfo  {journal} {Physical Review A}\ }\textbf {\bibinfo {volume} {45}},\ \bibinfo {pages} {R6149} (\bibinfo {year} {1992})}\BibitemShut {NoStop}%
\bibitem [{\citenamefont {Fahey}\ \emph {et~al.}(2023)\citenamefont {Fahey}, \citenamefont {Jacobs}, \citenamefont {Turner}, \citenamefont {Choi}, \citenamefont {Hoffman}, \citenamefont {Englund},\ and\ \citenamefont {Trusheim}}]{fahey2023steady}%
  \BibitemOpen
  \bibfield  {author} {\bibinfo {author} {\bibfnamefont {D.~P.}\ \bibnamefont {Fahey}}, \bibinfo {author} {\bibfnamefont {K.}~\bibnamefont {Jacobs}}, \bibinfo {author} {\bibfnamefont {M.~J.}\ \bibnamefont {Turner}}, \bibinfo {author} {\bibfnamefont {H.}~\bibnamefont {Choi}}, \bibinfo {author} {\bibfnamefont {J.~E.}\ \bibnamefont {Hoffman}}, \bibinfo {author} {\bibfnamefont {D.}~\bibnamefont {Englund}},\ and\ \bibinfo {author} {\bibfnamefont {M.~E.}\ \bibnamefont {Trusheim}},\ }\bibfield  {title} {\bibinfo {title} {Steady-state microwave mode cooling with a diamond n-v ensemble},\ }\href@noop {} {\bibfield  {journal} {\bibinfo  {journal} {Physical Review Applied}\ }\textbf {\bibinfo {volume} {20}},\ \bibinfo {pages} {014033} (\bibinfo {year} {2023})}\BibitemShut {NoStop}%
\bibitem [{\citenamefont {Huillery}\ \emph {et~al.}(2021)\citenamefont {Huillery}, \citenamefont {Leibold}, \citenamefont {Delord}, \citenamefont {Nicolas}, \citenamefont {Achard}, \citenamefont {Tallaire},\ and\ \citenamefont {H{\'e}tet}}]{huillery2021coherent}%
  \BibitemOpen
  \bibfield  {author} {\bibinfo {author} {\bibfnamefont {P.}~\bibnamefont {Huillery}}, \bibinfo {author} {\bibfnamefont {J.}~\bibnamefont {Leibold}}, \bibinfo {author} {\bibfnamefont {T.}~\bibnamefont {Delord}}, \bibinfo {author} {\bibfnamefont {L.}~\bibnamefont {Nicolas}}, \bibinfo {author} {\bibfnamefont {J.}~\bibnamefont {Achard}}, \bibinfo {author} {\bibfnamefont {A.}~\bibnamefont {Tallaire}},\ and\ \bibinfo {author} {\bibfnamefont {G.}~\bibnamefont {H{\'e}tet}},\ }\bibfield  {title} {\bibinfo {title} {Coherent microwave control of a nuclear spin ensemble at room temperature},\ }\href@noop {} {\bibfield  {journal} {\bibinfo  {journal} {Physical Review B}\ }\textbf {\bibinfo {volume} {103}},\ \bibinfo {pages} {L140102} (\bibinfo {year} {2021})}\BibitemShut {NoStop}%
\bibitem [{\citenamefont {Jamonneau}\ \emph {et~al.}(2016)\citenamefont {Jamonneau}, \citenamefont {H{\'e}tet}, \citenamefont {Dr{\'e}au}, \citenamefont {Roch},\ and\ \citenamefont {Jacques}}]{jamonneau2016coherent}%
  \BibitemOpen
  \bibfield  {author} {\bibinfo {author} {\bibfnamefont {P.}~\bibnamefont {Jamonneau}}, \bibinfo {author} {\bibfnamefont {G.}~\bibnamefont {H{\'e}tet}}, \bibinfo {author} {\bibfnamefont {A.}~\bibnamefont {Dr{\'e}au}}, \bibinfo {author} {\bibfnamefont {J.-F.}\ \bibnamefont {Roch}},\ and\ \bibinfo {author} {\bibfnamefont {V.}~\bibnamefont {Jacques}},\ }\bibfield  {title} {\bibinfo {title} {Coherent population trapping of a single nuclear spin under ambient conditions},\ }\href@noop {} {\bibfield  {journal} {\bibinfo  {journal} {Physical review letters}\ }\textbf {\bibinfo {volume} {116}},\ \bibinfo {pages} {043603} (\bibinfo {year} {2016})}\BibitemShut {NoStop}%
\bibitem [{\citenamefont {Vanier}(2005)}]{vanier2005atomic}%
  \BibitemOpen
  \bibfield  {author} {\bibinfo {author} {\bibfnamefont {J.}~\bibnamefont {Vanier}},\ }\bibfield  {title} {\bibinfo {title} {Atomic clocks based on coherent population trapping: a review},\ }\href@noop {} {\bibfield  {journal} {\bibinfo  {journal} {Applied Physics B}\ }\textbf {\bibinfo {volume} {81}},\ \bibinfo {pages} {421} (\bibinfo {year} {2005})}\BibitemShut {NoStop}%
\bibitem [{\citenamefont {Gray}\ \emph {et~al.}(1978)\citenamefont {Gray}, \citenamefont {Whitley},\ and\ \citenamefont {Stroud}}]{gray1978coherent}%
  \BibitemOpen
  \bibfield  {author} {\bibinfo {author} {\bibfnamefont {H.}~\bibnamefont {Gray}}, \bibinfo {author} {\bibfnamefont {R.}~\bibnamefont {Whitley}},\ and\ \bibinfo {author} {\bibfnamefont {C.}~\bibnamefont {Stroud}},\ }\bibfield  {title} {\bibinfo {title} {Coherent trapping of atomic populations},\ }\href@noop {} {\bibfield  {journal} {\bibinfo  {journal} {Optics letters}\ }\textbf {\bibinfo {volume} {3}},\ \bibinfo {pages} {218} (\bibinfo {year} {1978})}\BibitemShut {NoStop}%
\bibitem [{Sup()}]{Supplementary}%
  \BibitemOpen
  \href@noop {} {}\bibinfo {note} {See Supplementary Information for additional details.}\BibitemShut {Stop}%
\bibitem [{\citenamefont {Cohen-Tannoudji}(1996)}]{cohen1996autler}%
  \BibitemOpen
  \bibfield  {author} {\bibinfo {author} {\bibfnamefont {C.~N.}\ \bibnamefont {Cohen-Tannoudji}},\ }\bibfield  {title} {\bibinfo {title} {The autler-townes effect revisited},\ }in\ \href@noop {} {\emph {\bibinfo {booktitle} {Amazing Light: A Volume Dedicated To Charles Hard Townes On His 80th Birthday}}}\ (\bibinfo  {publisher} {Springer},\ \bibinfo {year} {1996})\ pp.\ \bibinfo {pages} {109--123}\BibitemShut {NoStop}%
\bibitem [{\citenamefont {Zibrov}\ \emph {et~al.}(1995)\citenamefont {Zibrov}, \citenamefont {Lukin}, \citenamefont {Nikonov}, \citenamefont {Hollberg}, \citenamefont {Scully}, \citenamefont {Velichansky},\ and\ \citenamefont {Robinson}}]{zibrov1995experimental}%
  \BibitemOpen
  \bibfield  {author} {\bibinfo {author} {\bibfnamefont {A.}~\bibnamefont {Zibrov}}, \bibinfo {author} {\bibfnamefont {M.}~\bibnamefont {Lukin}}, \bibinfo {author} {\bibfnamefont {D.}~\bibnamefont {Nikonov}}, \bibinfo {author} {\bibfnamefont {L.}~\bibnamefont {Hollberg}}, \bibinfo {author} {\bibfnamefont {M.}~\bibnamefont {Scully}}, \bibinfo {author} {\bibfnamefont {V.}~\bibnamefont {Velichansky}},\ and\ \bibinfo {author} {\bibfnamefont {H.}~\bibnamefont {Robinson}},\ }\bibfield  {title} {\bibinfo {title} {Experimental demonstration of laser oscillation without population inversion via quantum interference in rb},\ }\href@noop {} {\bibfield  {journal} {\bibinfo  {journal} {Physical Review Letters}\ }\textbf {\bibinfo {volume} {75}},\ \bibinfo {pages} {1499} (\bibinfo {year} {1995})}\BibitemShut {NoStop}%
\bibitem [{\citenamefont {Mompart}\ and\ \citenamefont {Corbalan}(2000)}]{mompart2000lasing}%
  \BibitemOpen
  \bibfield  {author} {\bibinfo {author} {\bibfnamefont {J.}~\bibnamefont {Mompart}}\ and\ \bibinfo {author} {\bibfnamefont {R.}~\bibnamefont {Corbalan}},\ }\bibfield  {title} {\bibinfo {title} {Lasing without inversion},\ }\href@noop {} {\bibfield  {journal} {\bibinfo  {journal} {Journal of Optics B: Quantum and Semiclassical Optics}\ }\textbf {\bibinfo {volume} {2}},\ \bibinfo {pages} {R7} (\bibinfo {year} {2000})}\BibitemShut {NoStop}%
\bibitem [{\citenamefont {Barry}\ \emph {et~al.}(2023)\citenamefont {Barry}, \citenamefont {Irion}, \citenamefont {Steinecker}, \citenamefont {Freeman}, \citenamefont {Kedziora}, \citenamefont {Wilcox},\ and\ \citenamefont {Braje}}]{barry2023ferrimagnetic}%
  \BibitemOpen
  \bibfield  {author} {\bibinfo {author} {\bibfnamefont {J.~F.}\ \bibnamefont {Barry}}, \bibinfo {author} {\bibfnamefont {R.~A.}\ \bibnamefont {Irion}}, \bibinfo {author} {\bibfnamefont {M.~H.}\ \bibnamefont {Steinecker}}, \bibinfo {author} {\bibfnamefont {D.~K.}\ \bibnamefont {Freeman}}, \bibinfo {author} {\bibfnamefont {J.~J.}\ \bibnamefont {Kedziora}}, \bibinfo {author} {\bibfnamefont {R.~G.}\ \bibnamefont {Wilcox}},\ and\ \bibinfo {author} {\bibfnamefont {D.~A.}\ \bibnamefont {Braje}},\ }\bibfield  {title} {\bibinfo {title} {Ferrimagnetic oscillator magnetometer},\ }\href@noop {} {\bibfield  {journal} {\bibinfo  {journal} {Physical Review Applied}\ }\textbf {\bibinfo {volume} {19}},\ \bibinfo {pages} {044044} (\bibinfo {year} {2023})}\BibitemShut {NoStop}%
\bibitem [{\citenamefont {Angerer}\ \emph {et~al.}(2017)\citenamefont {Angerer}, \citenamefont {Putz}, \citenamefont {Krimer}, \citenamefont {Astner}, \citenamefont {Zens}, \citenamefont {Glattauer}, \citenamefont {Streltsov}, \citenamefont {Munro}, \citenamefont {Nemoto}, \citenamefont {Rotter} \emph {et~al.}}]{angerer2017ultralong}%
  \BibitemOpen
  \bibfield  {author} {\bibinfo {author} {\bibfnamefont {A.}~\bibnamefont {Angerer}}, \bibinfo {author} {\bibfnamefont {S.}~\bibnamefont {Putz}}, \bibinfo {author} {\bibfnamefont {D.~O.}\ \bibnamefont {Krimer}}, \bibinfo {author} {\bibfnamefont {T.}~\bibnamefont {Astner}}, \bibinfo {author} {\bibfnamefont {M.}~\bibnamefont {Zens}}, \bibinfo {author} {\bibfnamefont {R.}~\bibnamefont {Glattauer}}, \bibinfo {author} {\bibfnamefont {K.}~\bibnamefont {Streltsov}}, \bibinfo {author} {\bibfnamefont {W.~J.}\ \bibnamefont {Munro}}, \bibinfo {author} {\bibfnamefont {K.}~\bibnamefont {Nemoto}}, \bibinfo {author} {\bibfnamefont {S.}~\bibnamefont {Rotter}}, \emph {et~al.},\ }\bibfield  {title} {\bibinfo {title} {Ultralong relaxation times in bistable hybrid quantum systems},\ }\href@noop {} {\bibfield  {journal} {\bibinfo  {journal} {Science advances}\ }\textbf {\bibinfo {volume} {3}},\ \bibinfo {pages} {e1701626} (\bibinfo {year} {2017})}\BibitemShut {NoStop}%
\bibitem [{\citenamefont {Wei}\ \emph {et~al.}(2020)\citenamefont {Wei}, \citenamefont {Wu}, \citenamefont {Hsiao}, \citenamefont {Tsai},\ and\ \citenamefont {Chen}}]{wei2020broadband}%
  \BibitemOpen
  \bibfield  {author} {\bibinfo {author} {\bibfnamefont {Y.-C.}\ \bibnamefont {Wei}}, \bibinfo {author} {\bibfnamefont {B.-H.}\ \bibnamefont {Wu}}, \bibinfo {author} {\bibfnamefont {Y.-F.}\ \bibnamefont {Hsiao}}, \bibinfo {author} {\bibfnamefont {P.-J.}\ \bibnamefont {Tsai}},\ and\ \bibinfo {author} {\bibfnamefont {Y.-C.}\ \bibnamefont {Chen}},\ }\bibfield  {title} {\bibinfo {title} {Broadband coherent optical memory based on electromagnetically induced transparency},\ }\href@noop {} {\bibfield  {journal} {\bibinfo  {journal} {Physical Review A}\ }\textbf {\bibinfo {volume} {102}},\ \bibinfo {pages} {063720} (\bibinfo {year} {2020})}\BibitemShut {NoStop}%
\bibitem [{\citenamefont {Trusheim}\ \emph {et~al.}(2020)\citenamefont {Trusheim}, \citenamefont {Jacobs}, \citenamefont {Hoffman}, \citenamefont {Fahey}, \citenamefont {Braje},\ and\ \citenamefont {Englund}}]{trusheim2020polariton}%
  \BibitemOpen
  \bibfield  {author} {\bibinfo {author} {\bibfnamefont {M.~E.}\ \bibnamefont {Trusheim}}, \bibinfo {author} {\bibfnamefont {K.}~\bibnamefont {Jacobs}}, \bibinfo {author} {\bibfnamefont {J.~E.}\ \bibnamefont {Hoffman}}, \bibinfo {author} {\bibfnamefont {D.~P.}\ \bibnamefont {Fahey}}, \bibinfo {author} {\bibfnamefont {D.~A.}\ \bibnamefont {Braje}},\ and\ \bibinfo {author} {\bibfnamefont {D.~R.}\ \bibnamefont {Englund}},\ }\bibfield  {title} {\bibinfo {title} {A polariton-stabilized spin clock},\ }\href@noop {} {\bibfield  {journal} {\bibinfo  {journal} {arXiv preprint arXiv:2009.02427}\ } (\bibinfo {year} {2020})}\BibitemShut {NoStop}%
\end{thebibliography}%

\end{document}